\documentclass[twoside,a4paper,11pt]{sea10}
\usepackage{graphicx}
\usepackage{hyperref}
\usepackage{movie15}
\usepackage{color}
\begin{document}
\pagenumbering{arabic}
\pagestyle{myheadings}
\thispagestyle{empty}
{\flushleft\includegraphics[width=\textwidth,bb=58 650 590 680]{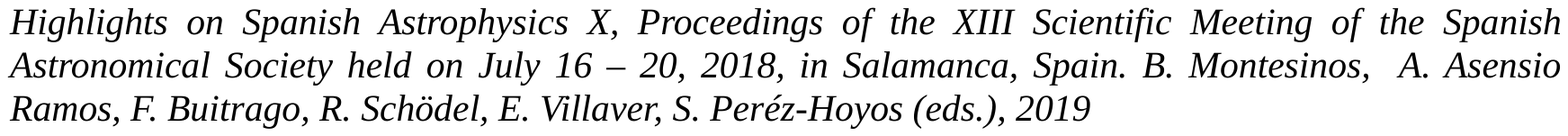}}
\vspace*{0.2cm}
\begin{flushleft}
{\bf {\LARGE
%
The Gaia photometric calibration and results on Galactic runaways
%
}\\
\vspace*{1cm}
%
Jes\'us Ma{\'\i}z Apell\'aniz$^1$, 
Michelangelo Pantaleoni Gonz\'alez$^{1,2}$,
Daniel J. Lennon$^{3}$, 
Rodolfo H. Barb\'a$^4$,
and 
Michael Weiler$^5$
%
}\\
\vspace*{0.5cm}
%
$^1$ 
Centro de Astrobiolog{\'\i}a, CSIC-INTA, Spain\\
$^2$ 
Universidad Complutense de Madrid, Spain\\
$^3$ 
European Space Agency, ESAC, Spain\\
$^4$ 
Universidad de La Serena, Chile\\
$^5$ 
Universitat de Barcelona, Spain\\
%
\end{flushleft}
%
\markboth{
The Gaia photometric calibration and results on Galactic runaways
}{ 
%
Ma{\'\i}z Apell\'aniz et al.
%
}
\thispagestyle{empty}
\vspace*{0.4cm}
\begin{minipage}[l]{0.09\textwidth}
\ 
\end{minipage}
\begin{minipage}[r]{0.9\textwidth}
\vspace{1cm}
\section*{Abstract}{\small
%
We present results on two different Gaia-related topics. First, we describe our efforts to calibrate the three Gaia photometric
passbands $G$, $G_{\rm BP}$, and $G_{\rm RP}$. We have built a new spectrophotometric HST/STIS library and used it to derive
new sensitivity curves and zero points for the three bands, including recipes on how to correct some cases. Second, we present our 
results on Galactic runaway stars using Gaia DR1 proper motions: we detect 76 runaway stars, 17 (possibly 19) of them not 
previously identified as such.
%
\normalsize}
\end{minipage}
%
%
%
\section{Introduction}

$\,\!$\indent In this contribution we present two different lines of work, both related to Gaia. First, we describe how we have
produced new passbands and zero points for the Gaia DR2 photometric system \cite{MaizWeil18}. Second, we describe the analysis of 
Gaia DR1 proper motions we have applied to detect runaway stars \cite{Maizetal18b}. 

\section{The Gaia photometric calibration}

$\,\!$\indent To fully compare spectral energy distribution (SED) models with photometric data one needs a passband curve and a 
zero point for each filter. Any errors in the passband lead to color (and possibly magnitude) corrections while errors in the zero 
point lead to photometric offsets. The passband curves for the Gaia photometric bands have been shown to differ from the 
laboratory measurements of \cite{Jordetal10}. For the case of DR1 $G$ photometry \cite{Maiz17a} and \cite{Weiletal18} provided 
updated curves. Later on, \cite{Evanetal18} and \cite{Weil18} provided updated curves for DR2 $G$+$G_{\rm BP}$+$G_{\rm RP}$ 
photometry but they are not fully satisfactory, as they cannot completely reproduce the observations (especially for $G_{\rm RP}$): 
a new study is needed.

\begin{figure}
\centerline{\includegraphics[width=0.490\linewidth]{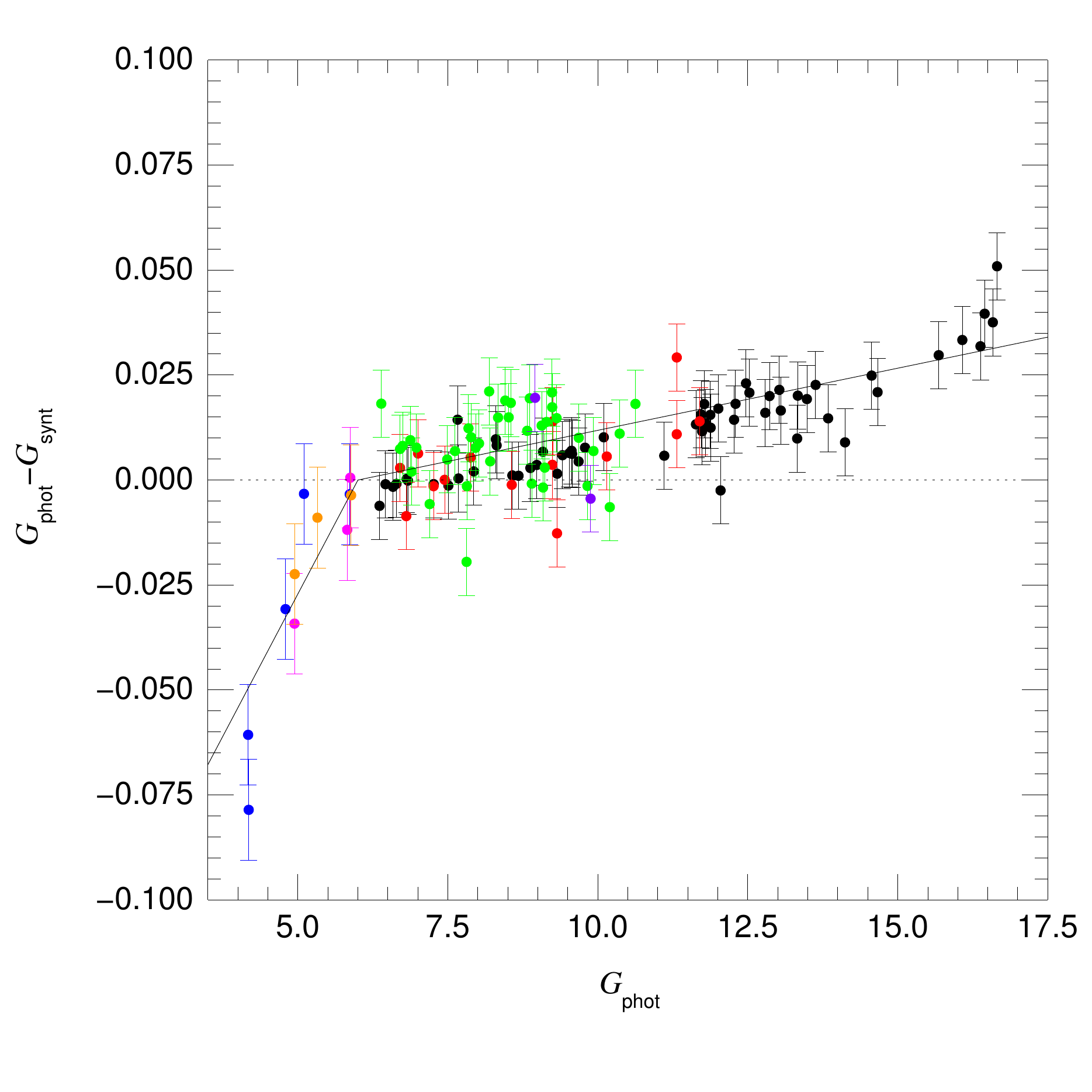} \
            \includegraphics[width=0.490\linewidth]{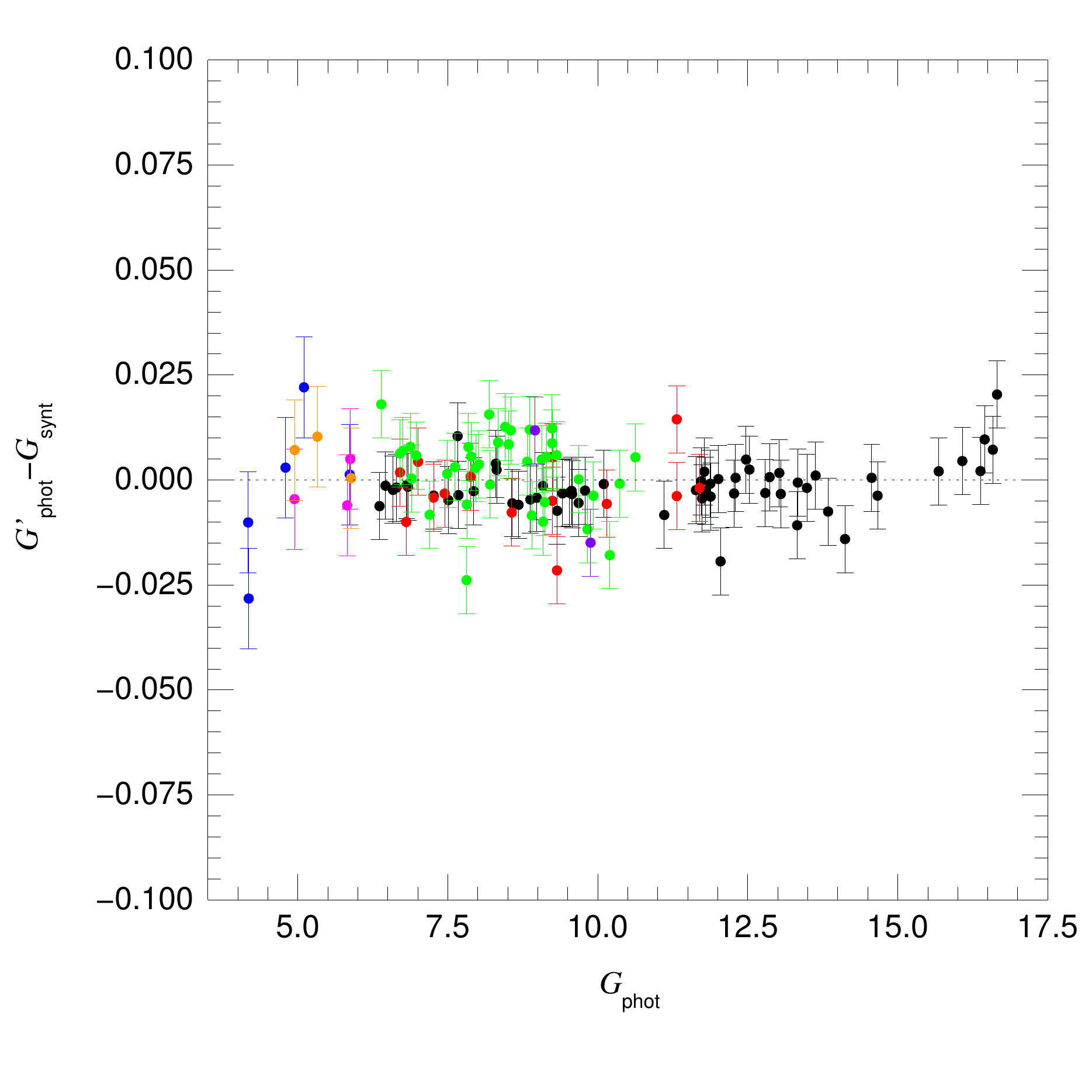}}
\caption{\label{fig1}Difference between observed and synthetic $G$ magnitudes as a function of $G$ before (left) and after (right) 
applying the correction. The vertical scale is the same in both plots. Colors are used to distinguish the data origins.}
\end{figure}

To generate new Gaia passband curves and zero points, we searched the STIS/HST archive to find stars with accurate Gaia DR2 
photometry observed with G430L+G750L and a wide slit. In the process we found and reduced a total of 122 objects: 63 from 
CALSPEC \cite{Bohletal17}, 17 from HOTSTAR \cite{KhanWort18}, 40 from an HST proposal by D. Massa \cite{Mass14}, and 2 from 
other projects. This new composite STIS library does not have the problems of previous SED libraries (ground-based or STIS with a 
narrow slit). It also includes three M dwarfs that are redder than other objects used to previously calibrate Gaia: 
(a) BD~$-$11~3759 = GJ~555 ($G_{\rm BP}-G_{\rm RP}$~=~2.932), (b) Proxima~Centauri = $\alpha$~Cen~C 
($G_{\rm BP}-G_{\rm RP}$~=~3.797), and (c) 2MASS~J16553529$-$0823401 = GJ~644~C = VB~8 ($G_{\rm BP}-G_{\rm RP}$~=~4.754).

Armed with the new STIS library, we have used it to recalibrate the Gaia DR2 photometry, generating new passbands and zero points.
Our results show that the quality of Gaia photometry is excellent but that corrections are needed.

\begin{figure}
\begin{minipage}{0.50\textwidth}
\centerline{\includegraphics[width=\linewidth]{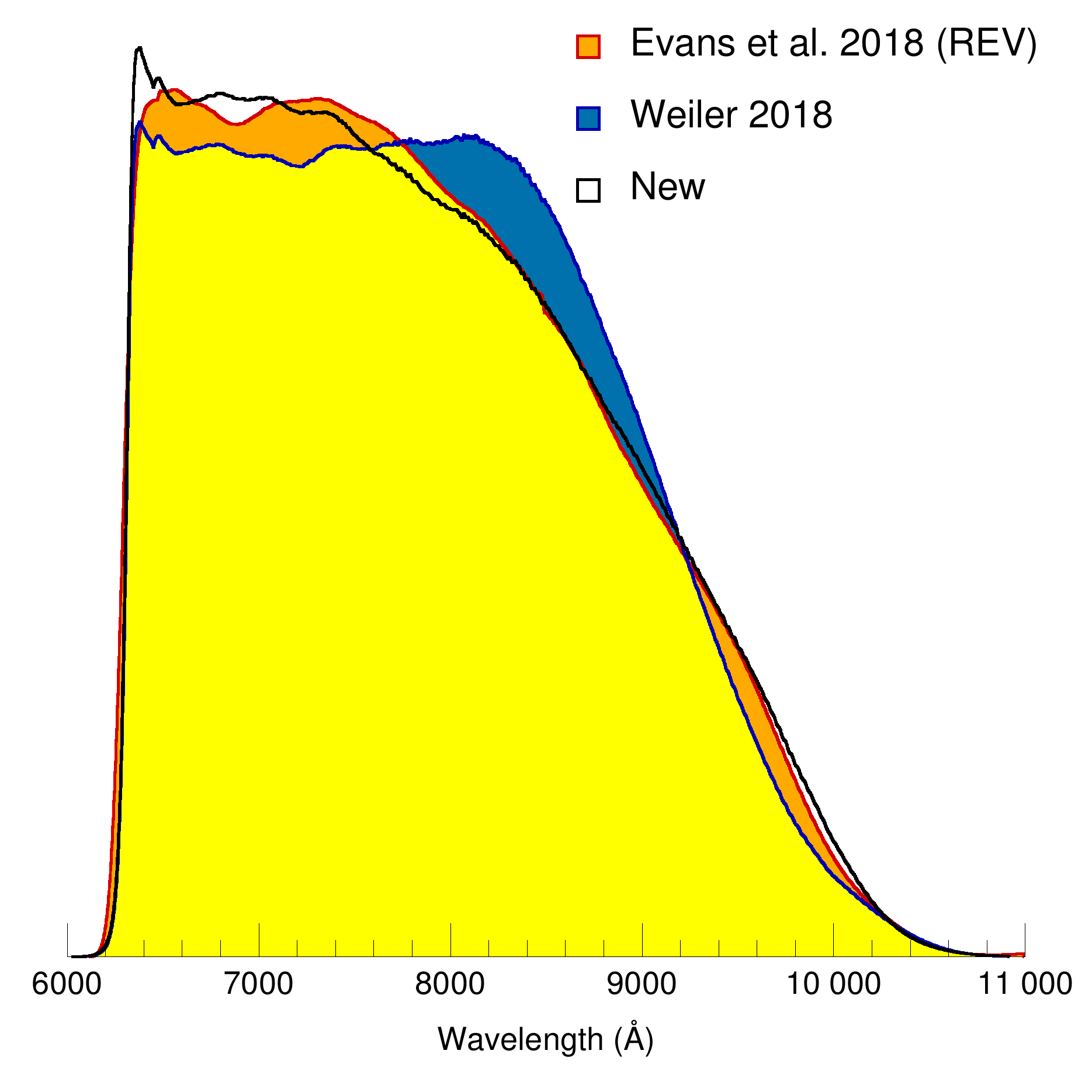}}
\end{minipage}
\begin{minipage}{0.48\textwidth}
\caption{\label{fig2}$G_{\rm RP}$ sensitivity curves from Evans et al. (2018), Weiler (2018), and this work. The three curves have 
been normalized to have the same area.}
\end{minipage}
\end{figure}

\begin{itemize}
 \item $G$ photometry has a very low dispersion (8 mmag) but requires a magnitude-dependent correction (Fig.~\ref{fig1}):
       $G^\prime = G - 0.0032\,(G - 6)$ for $G > 6$ and $G^\prime = G + 0.0271\,(G - 6)$ for $G < 6$.
 \item $G_{\rm BP}$ photometry has a slightly higher dispersion (9 mmag) and requires two different sensitivity curves, one for
       $G < 10.87$ and another for $G > 10.87$. A significant discontinuity is detected at that magnitude.
 \item $G_{\rm RP}$ photometry has a dispersion of 10 mmag for most objects and was previously poorly characterized for very red 
       objects (Fig.~\ref{fig2}). In any case, it will be necessary to increase the spectrophotometric sample for objects with 
       $G_{\rm BP}-G_{\rm RP} >$~1.5 in order to improve the accuracy for such red objects.
\end{itemize}

The new passbands and zero points are more accurate than the previous attempts by \cite{Evanetal18} and \cite{Weil18}. More details
are found in \cite{Maizetal18b}.

\section{Finding runaway stars in Gaia DR1}

$\,\!$\indent We have used Gaia~DR1 proper motions to find new runaway stars. A preliminary version of this work appeared
in \cite{Maizetal17d} and more details are given in \cite{Maizetal18b}. We used the Galactic O-Star Catalog (GOSC,
\url{http://gosc.cab.inta-csic.es}, \cite{Maizetal04b,Maizetal12,Maizetal17c,Sotaetal08}) to select two samples:
(a) objects with a Galactic O-Star Spectroscopic Survey (GOSSS, \cite{Maizetal11,Maizetal16,Sotaetal11a,Sotaetal14}) spectral 
classifications as O and (b) other Simbad O stars and BA supergiants for which we obtained a GOSSS spectrum, had reliable 
classifications, or had their intrinsic colors confirmed by a CHORIZOS analysis \cite{Maiz04c}. 

\begin{figure}
\begin{minipage}{0.65\textwidth}
\centerline{\includegraphics[width=\linewidth]{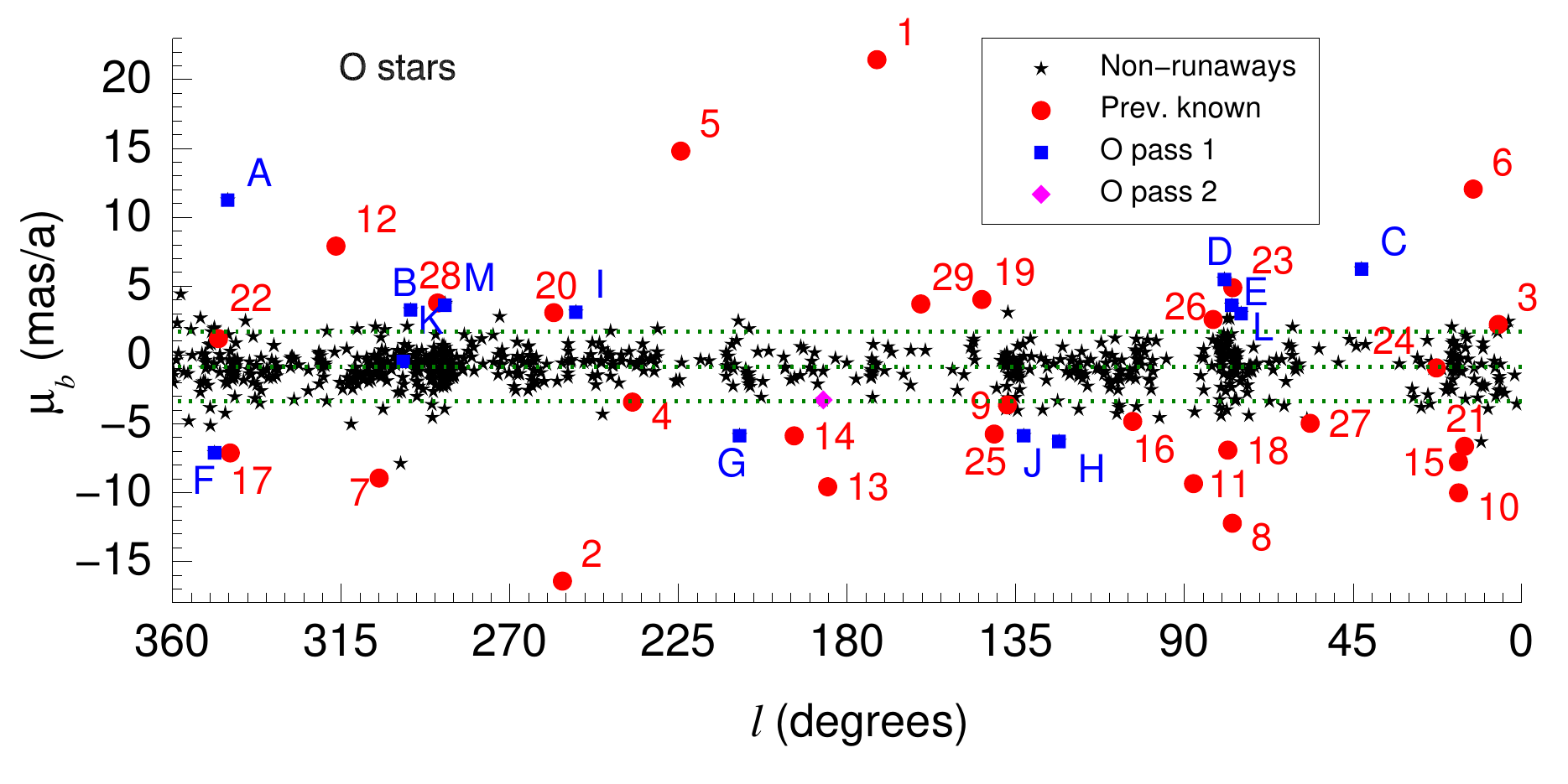}}
\centerline{\includegraphics[width=\linewidth]{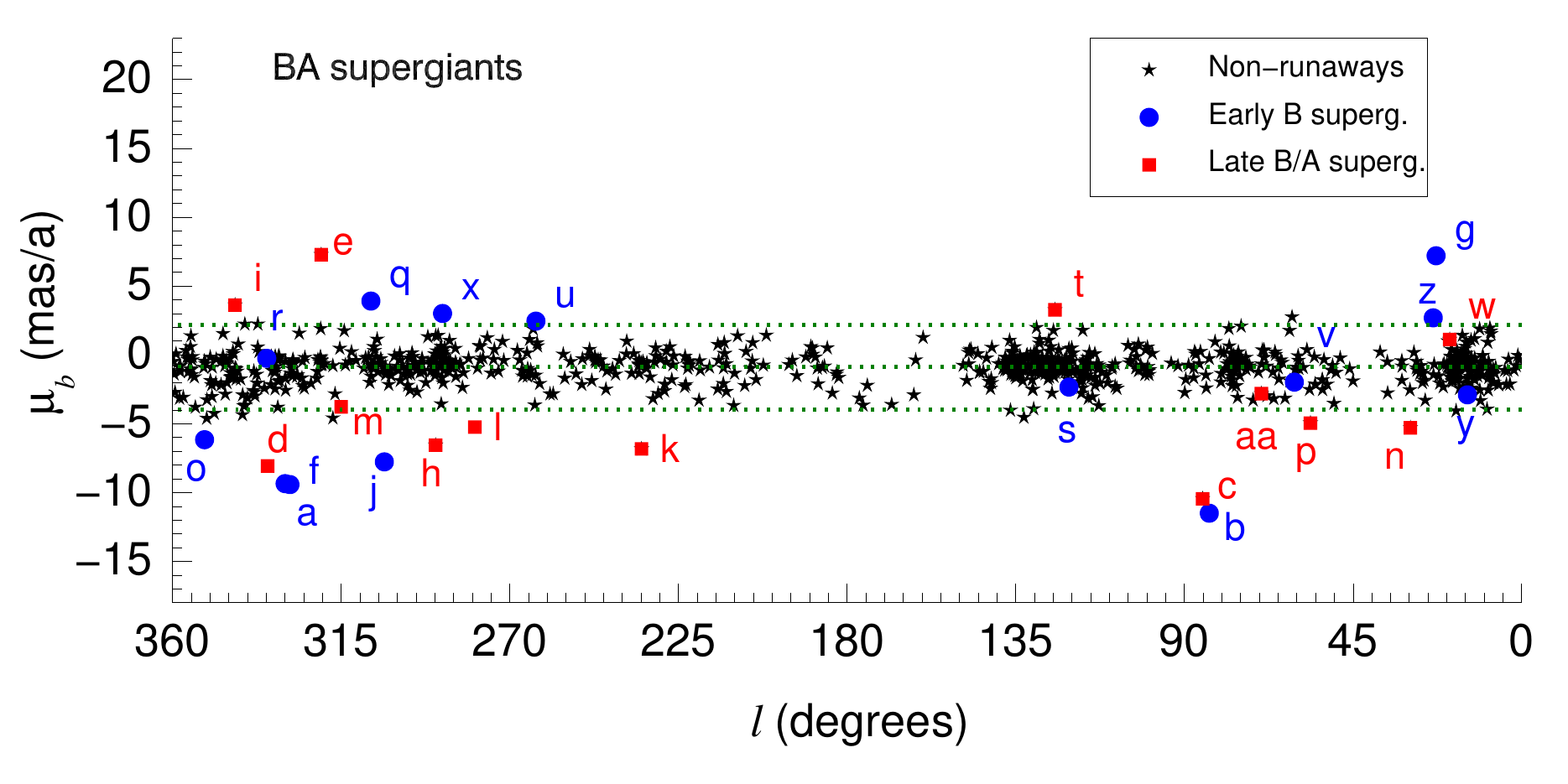}}
\end{minipage}
\begin{minipage}{0.35\textwidth}
\caption{\label{fig3}Observed proper motion in Galactic latitude for O stars (top) and BA supergiants (bottom). Different colors and
symbols are used to identify the runaways candidates (see \cite{Maizetal18b} for details). The dotted green lines represent the 
functions and 2$\sigma$ deviations used to detect runaway candidates.}
\end{minipage}
\end{figure}

\begin{figure}
\begin{minipage}{0.65\textwidth}
\centerline{\includegraphics[width=\linewidth]{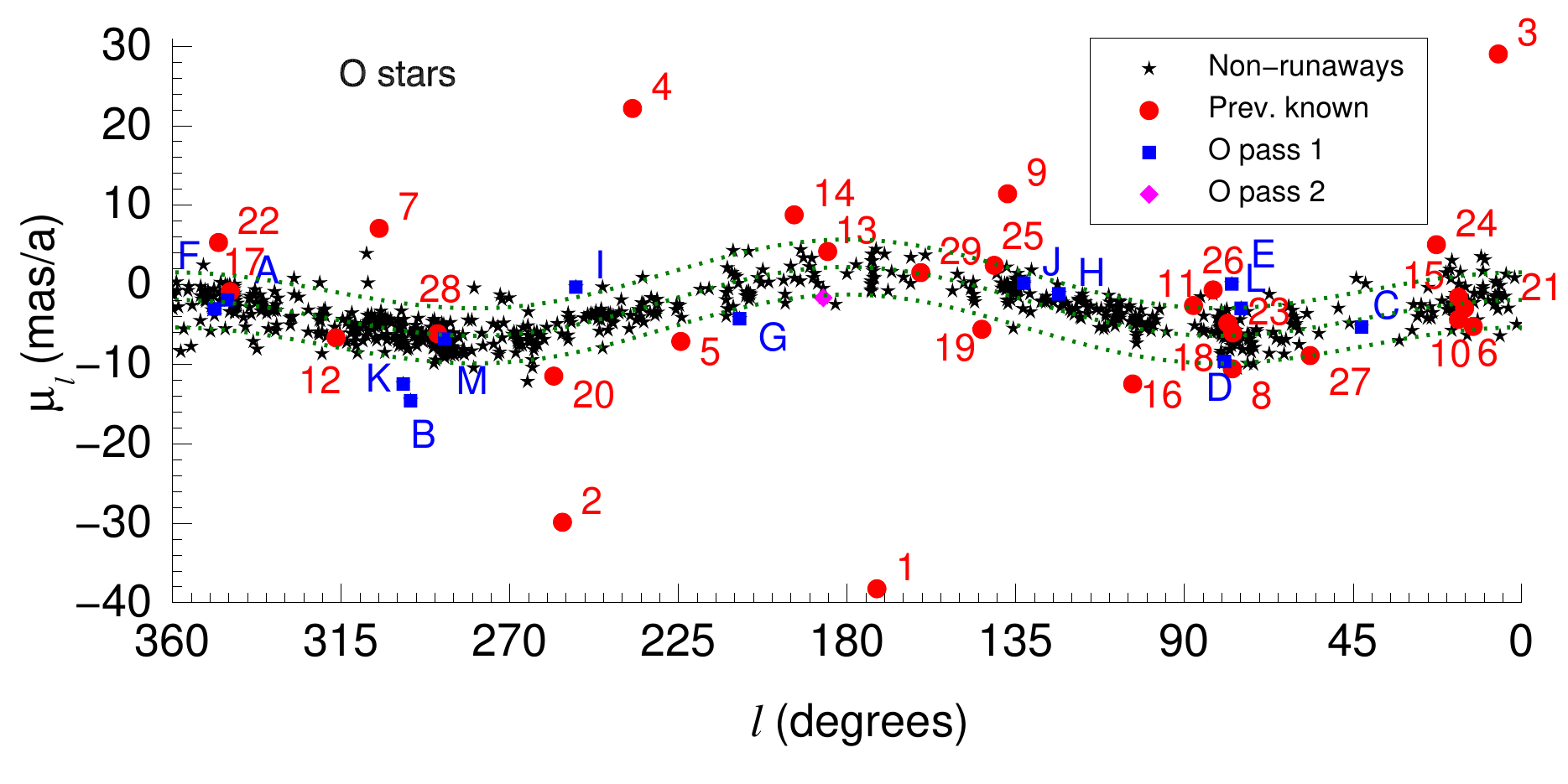}}
\centerline{\includegraphics[width=\linewidth]{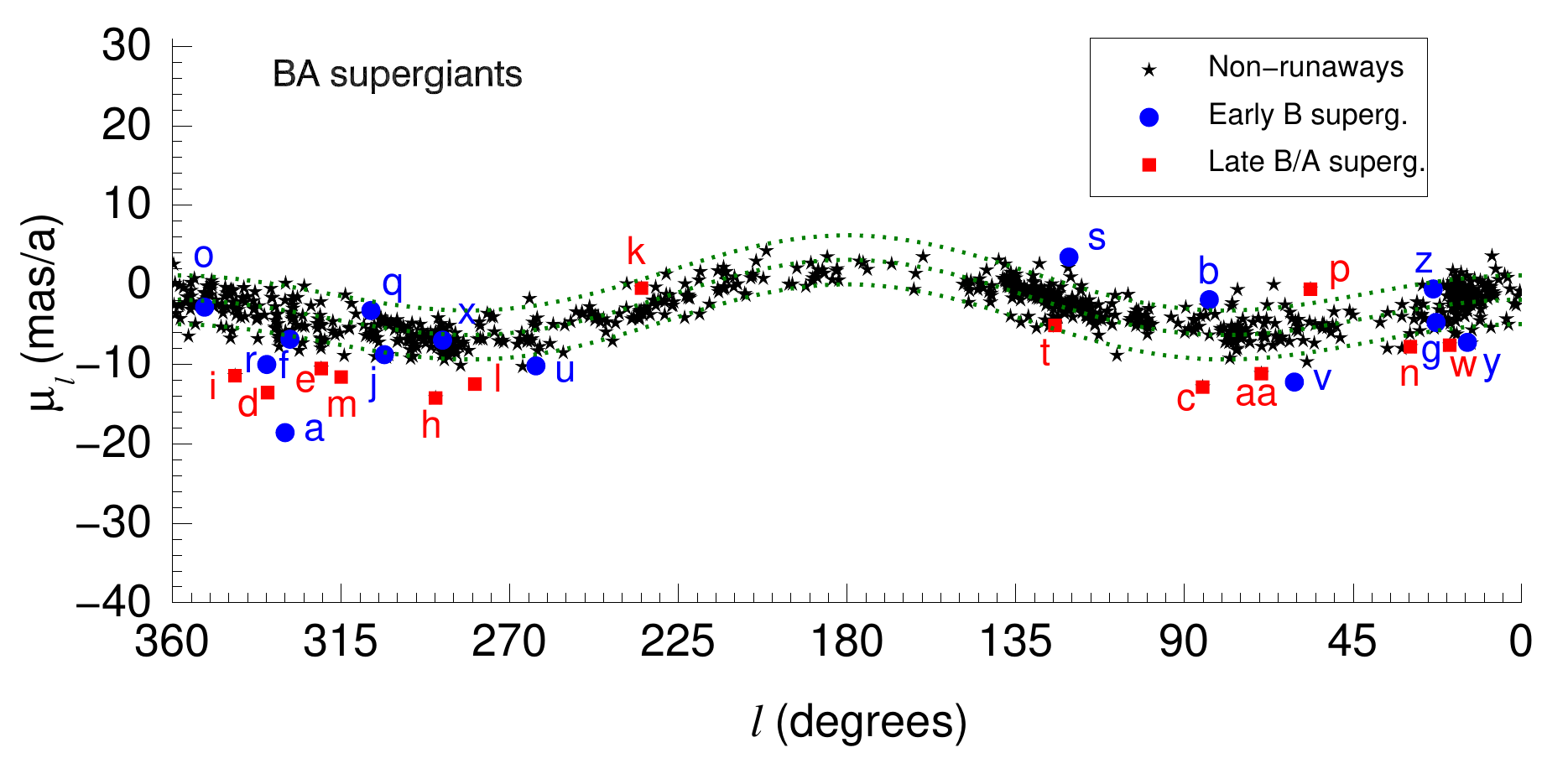}}
\end{minipage}
\begin{minipage}{0.35\textwidth}
\caption{\label{fig4}Observed proper motion in Galactic longitude for O stars (top) and BA supergiants (bottom). Different colors 
and symbols are used to identify the runaways candidates (see \cite{Maizetal18b} for details). The dotted green lines represent the 
functions and 2$\sigma$ deviations used to detect runaway candidates.}
\end{minipage}
\end{figure}

\begin{figure}
\centerline{\hspace{0.09\textwidth} \
            \includegraphics[width=0.490\linewidth, bb=52 28 541 725]{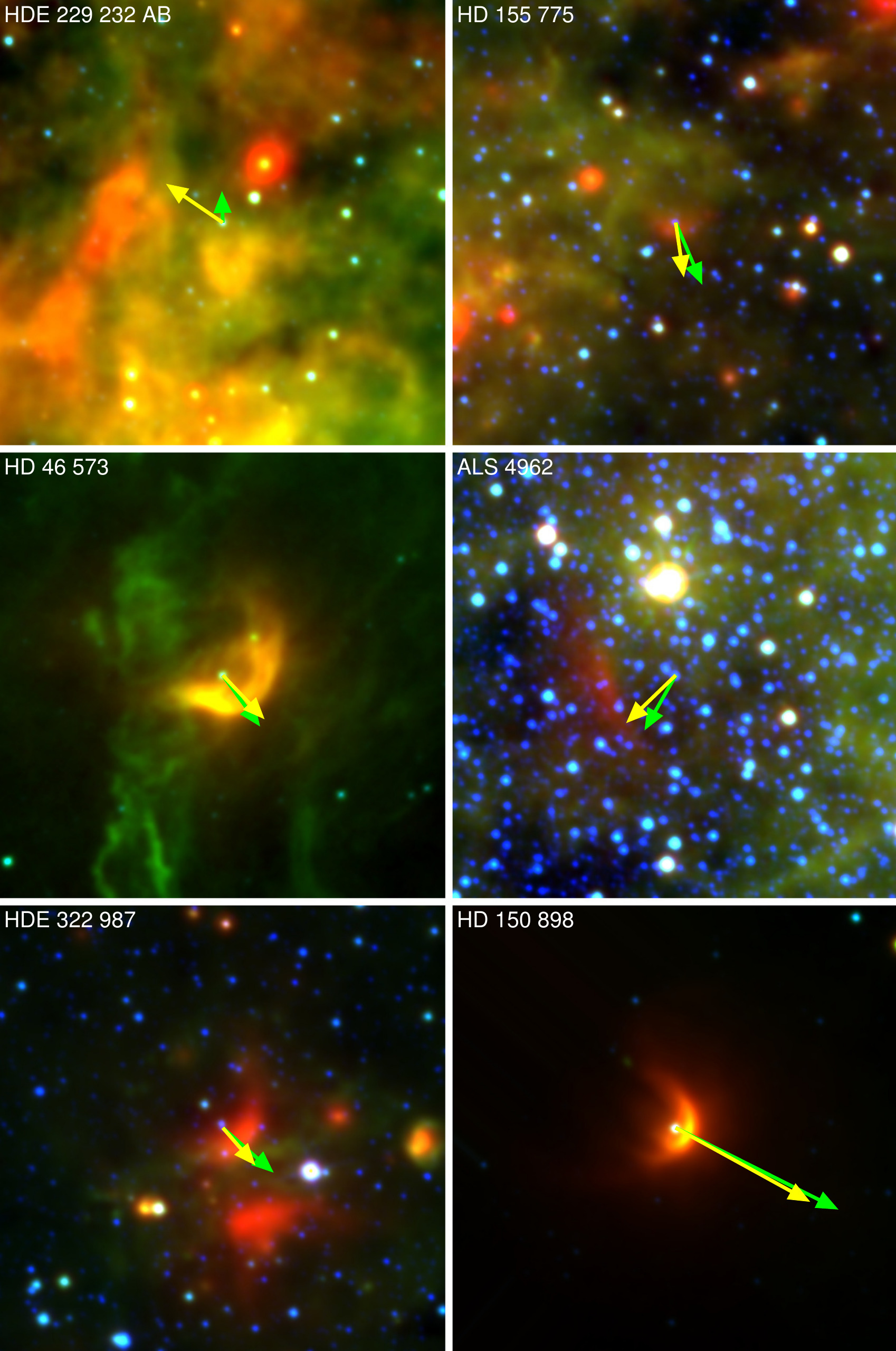} \
            \includegraphics[width=0.490\linewidth, bb=52 28 541 725]{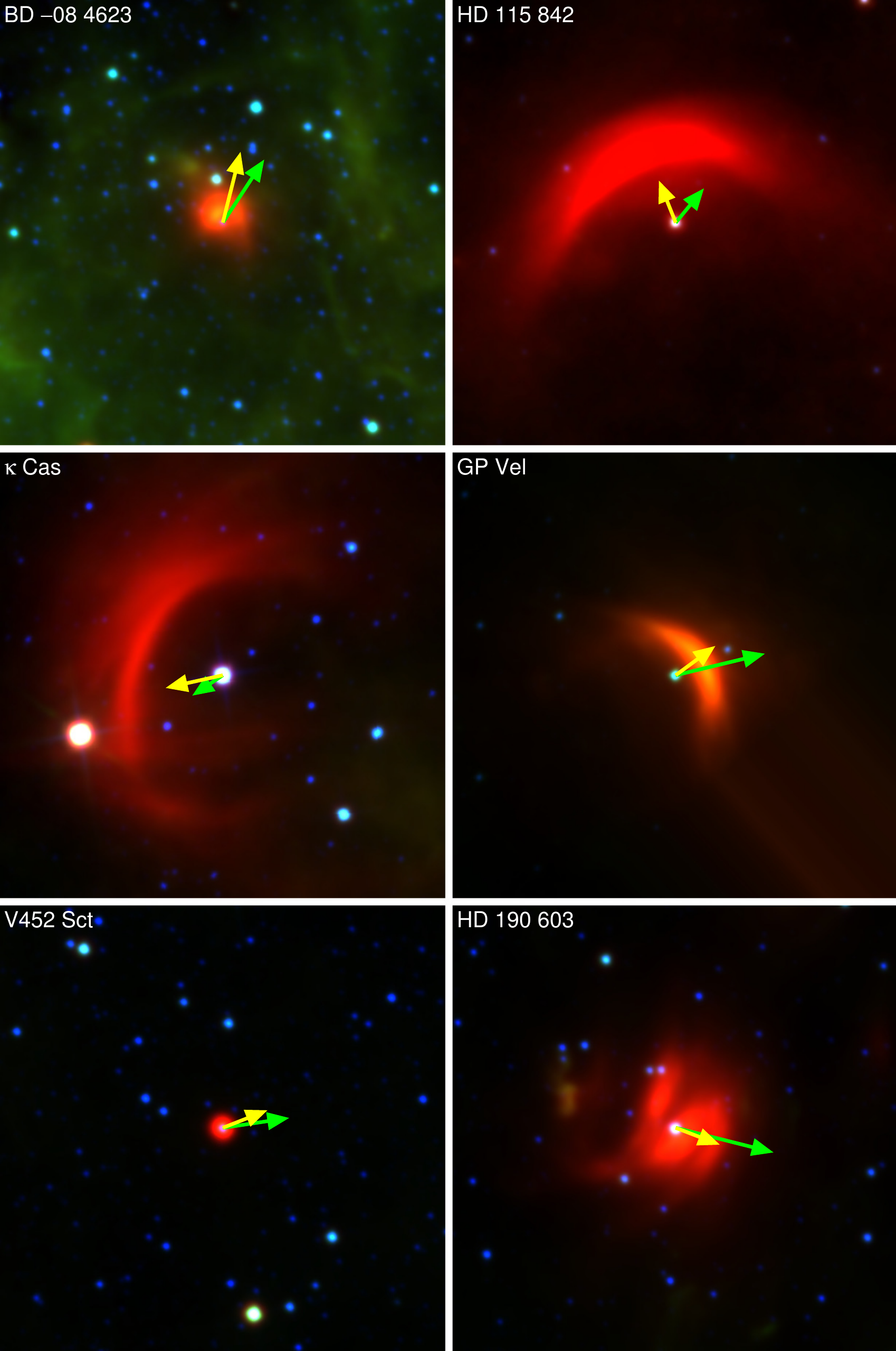}}
\caption{\label{fig5}WISE W4+W3+W2 RGB mosaics for twelve runaway candidates. Each field is 14$^\prime\times$14$^\prime$ and is 
oriented with Galactic North towards the top and Galactic East towards the left. In each mosaic the runaway candidate is at the 
center and the arrows show the original proper motion (green) and the corrected one (yellow).}
\end{figure}

For those two samples we obtained the TGAS (Gaia DR1) proper 
motions complemented with Hipparcos for bright stars. We then selected the runaway stars by fitting longitude and latitude proper 
motion values as a function of longitude, calculating the corrected proper motions, and finding the outliers 
(Figs.~\ref{fig3}~and~\ref{fig4}). We identified 76 runaway stars, 17 to 19 for the first time. We also searched for bow shocks in
the MIR using WISE images and we found them for some objects (Figs.~\ref{fig5})

%
%
\small  
%
\section*{Acknowledgments}   
%
J.M.A. acknowledges support from the Spanish Government Ministerio de Ciencia, Innovaci\'on y Universidades through grant
AYA2016-75\,931-C2-2-P. 
M.P.G. acknowledges support from the ESAC Trainee program.
R.H.B. acknowledges support from the ESAC Faculty Council Visitor Program. 
M.W. acknowledges support from the Spanish Government Ministerio de Ciencia, Innovaci\'on y Universidades through grants 
ESP2016-80\,079-C2-1-R (MINECO/FEDER, UE) and ESP2014-55996-C2-1-R (MINECO/FEDER, UE) and MDM-2014-0369 of ICCUB (Unidad de Excelencia ''Mar{\'\i}a de Maeztu''.
%

%
\end{document}